\documentstyle[12pt]{article}

\newcommand\T{\theta_{12}}
\newcommand\Tb{{\bar\theta}_{12}}
\newcommand\Z{Z_{12}}
\newcommand\D{{\cal D}}
\newcommand\Db{\overline{\cal D}}
\newcommand\jl{J(Z_1)}
\newcommand\jr{J(Z_2)}
\newcommand\tl{T(Z_1)}
\newcommand\tr{T(Z_2)}
\newcommand\Ub{{\bar{\cal U}}}
\newcommand\Pb{\bar{\Phi}}
\newcommand\U{{\cal U}}
\newcommand\Ph{\Phi}

\begin{document}
\renewcommand{\thefootnote}{\fnsymbol{footnote}}
\thispagestyle{empty}
{\hfill  IC/92/64 }\vspace{0.5cm} \\
\begin{center}
International Atomic Energy Agency \vspace{0.2cm}\\
and \vspace{0.2cm}\\
United Nations Educational Scientific and Cultural Organization
\vspace{0.3cm}\\
INTERNATIONAL CENTRE FOR THEORETICAL PHYSICS \vspace{3.5cm} \\
{\large\bf SUPERFIELD REALIZATIONS OF $N=2$
SUPER-$W_3$}
\vspace{1.5cm} \\
 E.Ivanov\footnote{Permanent address: Laboratory of Theoretical Physics,
JINR, Dubna, Russian Federation.}\footnote{E-mail address:
eivanov@ltp.jinr.dubna.su}
and S.Krivonos${}^\star$\footnote{E-mail address: krivonos@ltp.jinr.dubna.su}
\vspace{0.8cm}\\
International Centre for Theoretical Physics, Trieste, Italy \vspace{2cm} \\
{\bf Abstract}
\end{center}

We present a manifestly $N=2$ supersymmetric formulation of
$N=2$ super-$W_3$ algebra (its classical version) in terms of the
spin 1 and spin 2 supercurrents. Two closely related types of the Feigin-Fuchs
representation for these supercurrents are found: via two chiral
spin $\frac{1}{2}$ superfields generating  $N=2$ extended $U(1)$ Kac-Moody
algebras and via two free chiral spin 0 superfields. We also construct a
one-parameter family of $N=2$ super Boussinesq equations for which $N=2$
super-$W_3$ provides the second hamiltonian structure. \vspace{0.5cm} \\
\begin{center}
{\it Submitted to Phys. Lett. B}
\end{center}
\vfill
\begin{center}
MIRAMARE-TRIESTE \\
 April 1992
\end{center}
\setcounter{page}0
\renewcommand{\thefootnote}{\arabic{footnote}}
\setcounter{footnote}0
\newpage
\section{Introduction}
For the last two years there has been considerable interest in
supersymmetric extensions of Zamolodchikov's $W_N$
algebras (see, e.g. \cite{a10}-\cite{a14}).
Recently, $N=2$ super-$W_3$ algebra has been constructed, both on the
classical \cite{{a8},{a6}}
and full quantum \cite{aa} levels. It is generated by two $N=2$
supermultiplets of currents, with the conformal spins
$(1,\;\frac{3}{2},\;\frac{3}{2},\;2)$ and $(2,\;\frac{5}{2},
\;\frac{5}{2},\;3)$,
and exists at an arbitrary value of the central charge.

For setting up the conformal field theory associated with $N=2$ super-$W_3$
and studying representations of this algebra it is of importance to
know its free-field realizations. One more urgent problem
which could
have important implications in $N=2$ super-$W_3$ gravity and
the related matrix models consists in defining the general hamiltonian flow
on $N=2$ super-$W_3$ and finding out the generalized KdV-type hierarchy for
which this algebra produces the second hamiltonian structure.

In the present letter we address both these problems. We give two
field-theoretical realizations of $N=2$ super-$W_3$: via spin $\frac{1}{2}$
and spin 0 chiral $N=2$ supermultiplets. Both these realizations are of the
Feigin-Fuchs type and place no restrictions on the central charge. We also
construct the simplest nontrivial hamiltonian flow on $N=2$ super-$W_3$
yielding a $N=2$ superextension of the Boussinesq equation. This
superextension turns out to involve a free parameter and is reducible
in the bosonic sector to the Boussinesq equation only at the special value
of this parameter.

We use the language of $N=2$
superfields which radically simplifies computations and allows to present
the final results in a manifestly $N=2$ supersymmetric concise form. We
restrict our study here to the classical version of $N=2$ super-$W_3$
\cite{a6}. Extension to the full quantum $N=2$ super-$W_3$ \cite{aa}
seems to be more or less straightforward and will be presented
elsewhere.

\section{$N=2$ super-$W_3$ algebra in terms of $N=2$ supercurrents}
In this Section we present a supercurrent formulation of the classical
$N=2$ super-$W_3$ algebra \cite{a6}.

The basic currents of $N=2$ super-$W_3$, in accordance with their spin
content, can be naturally incorporated into the two
$N=2$ supercurrents $J(Z)$ and $T(Z)$ carrying the
spins 1 and 2.\footnote{By $Z$ we denote the coordinates of $1D$
$N=2$ superspace, $Z=(z,\theta,\bar\theta ).$}. The full set of the current
OPE's of ref.\cite{a6} can now be summarized as the three SOPE's between
these supercurrents.

The first SOPE states that the
supecurrent $J(Z)$ generates the standard $N=2$ superconformal
algebra \cite{a2}-\cite{a4}:
\begin{equation}\label{1}
\jl\jr=\frac{{c\over 4}}{\Z^2}+ \frac{\Tb\Db J}{\Z}-\frac{\T\D J}{\Z}
+\frac{\T\Tb J}{\Z^2}+\frac{\T\Tb\partial J}{\Z} \; ,
\end{equation}
where
\begin{equation}\label{2}
\T=\theta_1-\theta_2 \quad , \quad \Tb=\bar\theta_1-\bar\theta_2 \quad ,
 \quad \Z=z_1-z_2+\frac{1}{2}\left( \theta_1\bar\theta_2
-\theta_2\bar\theta_1 \right) \quad ,
\end{equation}
and $\D,\Db$ are the spinor covariant derivatives
\begin{equation}\label{3}
\D=\frac{\partial}{\partial\theta}
 -\frac{1}{2}\bar\theta\frac{\partial}{\partial z} \quad , \quad
\Db=\frac{\partial}{\partial\bar\theta}
 -\frac{1}{2}\theta\frac{\partial}{\partial z}
\end{equation}
$$
\left\{\D,\Db \right\}= -\frac{\partial}{\partial z} \quad , \quad
\left\{\D,\D \right\} = \left\{\Db,\Db \right\}= 0.
$$

The next SOPE expresses the property that the spin 2 supercurrent
$T(Z)$ can be chosen primary with
respect to the $N=2$ superconformal algebra:
\begin{equation}\label{4}
\jl\tr= \frac{\Tb\Db T}{\Z}-\frac{\T\D  T}{\Z}
+2\frac{\T\Tb T}{\Z^2}+\frac{\T\Tb\partial T}{\Z} \; .
\end{equation}
(it possesses zero $U(1)$ charge).

The last SOPE needed to close the algebra is that involving $\tl,\;\tr$. It
is most complicated
$$
\tl\tr =  -\frac{{3\over 2}c}{\Z^4}-12\frac{\T\Tb J}{\Z^4}+
12\frac{\T\D J}{\Z^3}-12\frac{\Tb\Db J}{\Z^3}-
12\frac{\T\Tb\partial J}{\Z^3}
$$
$$
+ 2\frac{5\tr -2\left[\Db,\D\right] J+{8\over c} J^2)}{\Z^2}
+ \frac{\T\D\left(  8\partial J + 5T
 +{8\over c}J^2\right)}{\Z^2}
$$
$$
-\frac{\Tb\Db\left(  8\partial J-5T
 -{8\over c}J^2\right)}{\Z^2}
 +  \frac{\T\Tb\left({3\over 2}\left[\Db,\D\right] T-
    6\partial^2 J+U_3\right)}{\Z^2}
$$
$$
 + \frac{\T\left( 3\partial\D T
 +3\partial^2\D J+\Psi_{{7\over 2}}
     \right)}{\Z}
    -\frac{\Tb\left(-3\partial\Db T+3\partial^2\Db J
    +{\overline\Psi}_{{7\over 2}}  \right)}{\Z}
$$
$$
+ \frac{\T\Tb}{\Z}\left( -2\partial^3 J+\partial\left[ \Db,\D\right] T
 + \frac{1}{2}\partial U_3 +\frac{1}{2}\Db\Psi_{{7\over 2}}
 + \frac{1}{2}\D{\overline\Psi}_{{7\over 2}}-
   \frac{2}{c}\partial \left[ \Db,\D \right] J^2 \right)
$$
\begin{equation}\label{5}
+  \frac{\partial\left( 5{\widetilde T}-2\left[ \Db,\D\right] J
   +{8\over c}J^2 \right)}{\Z} \quad .
\end{equation}
Here  $ \Psi_{{7\over 2}}(Z),\;{\overline\Psi}_{{7\over 2}}(Z),\;
U_3(Z) $ are the composite supercurrents of the spins
$ {7\over 2},\;{7\over 2},\;3 $,
respectively
\begin{eqnarray}
\Psi_{{7\over 2}} & = & \frac{8}{c} \partial\left( J\D J\right)
   -\frac{72}{c}T\D J +\frac{36}{c}\left[\Db ,\D\right]J \D J +
 \frac{8}{c}J\D T -\frac{128}{c^2}J^2\D J
	    +\frac{4}{c}\partial J\D J \nonumber \\
{\overline\Psi}_{{7\over 2}} & = & -\frac{8}{c} \partial\left( J\Db J\right)
   -\frac{72}{c} T\Db J +\frac{36}{c}\left[\Db ,\D\right]J \Db J +
\frac{8}{c}J\Db T -\frac{128}{c^2}J^2\Db J
	    -\frac{4}{c}\partial J\Db J \nonumber \\
U_3 & = & \frac{56}{c}J T -\frac{32}{c}J\left[ \Db ,\D \right] J
	 +\frac{128}{c^2}J^3 +\frac{120}{c}\Db J\D J \quad . \label{6}
\end{eqnarray}

The complete correspondence with ref. \cite{a6} is achieved under the
following definition of the component currents
\begin{eqnarray}
 J | = 4J_0 & , & T| =
 T_0+4\widetilde{T}_0-\frac{128}{c}J^2_0 \nonumber \\
 \D J | = \bar{G}_0 & , & \D T| =\frac{3}{4}\bar{U}_0-
           \frac{64}{c}J_0\bar{G}_0 \nonumber \\
\Db J | =-G_0 & , & \Db T| = -\frac{3}{4}U_0+
           \frac{64}{c}J_0 G_0 \label{7} \\
\frac{1}{2}\left[ \Db,\D\right]J | = T_0+\widetilde{T}_0 & , &
  \frac{1}{2}\left[ \Db,\D\right] T| =
 \frac{3}{4}W_0+\frac{32}{c}\left( T_0+4\widetilde{T}_0-\frac{128}{c}J^2_0
 \right) J_0 +\frac{40}{c}G_0\bar{G}_0 \;  . \nonumber
\end{eqnarray}
The currents $ J_0,\;G_0,\;\bar{G}_0,\;T_0,\;\widetilde{T}_0,\;U_0,
\;\bar{U}_0,\;
W_0 $ obey the OPE's of \cite{a6} as a consequence of
the SOPE's (\ref{1}), (\ref{4}), (\ref{5}).

It is worth noting that the spin content of our
composite N=2 supercurrents $ J^2,\; \Psi_{{7\over 2}},\;
\overline{\Psi}_{{7\over 2}},\;
U_3 $ is larger than that of the set of composite currents figuring
in the OPE's of ref. \cite{a6}:
$$
\left( 2,{5\over 2},{5\over 2},3;\; {7\over 2},{7\over 2},4,4,{9\over 2},
{9\over 2}; \; 3,{7\over 2},{7\over 2},4 \right) \quad \mbox{vs} \quad
\left( 3,{7\over 2},{7\over 2},4,4,{9\over 2},{9\over 2},
3,{7\over 2},{7\over 2} \right).
$$
This differences stems from the fact that we are working in a manifestly
$N=2$ supersymmetric superfield formalism, so the composites can appear only
in groups forming $N=2$ supermultiplets. Of course, the missing composite
currents are implicitly present also in the relations of ref. \cite{a6}:
they can be produced by action of the $N=2$ supersymmetry
generators on the composites appearing explicitly.

Thus we have established the full $N=2$ superfields structure of
$N=2$ super-$W_3$ algebra. Before closing this Section, let us indicate one
more version of classical $N=2$ super-$W_3$ which is similar to the algebra
called in ref. \cite{a13} ``classical $W_3$'' and follows from the $N=2$
super-$W_3$ defined above in the contraction limit $c=0$. We call this
contracted superalgebra $N=2$ super-$W_3^{cl}$.

In order to approach the limit $c\rightarrow 0$ in an
unambiguous way, one needs beforehand to rescale the  supercurrents
as
\begin{equation}\label{8}
J^{cl}=J\quad , \quad
T^{cl}=cT\quad ,\quad
\Psi_{{7\over 2}}^{cl} = c^2\Psi_{{7\over 2}}\quad , \quad
{\overline\Psi}_{{7\over 2}}^{cl} =c^2{\overline\Psi}_{{7\over 2}}\quad , \quad
U_{3}^{cl} = c^2U_{3} \;.
\end{equation}
Now it is straightforward to put $c=0$ in SOPE's (\ref{1}), (\ref{4}),
(\ref{5}) and to obtain the algebra $N=2$ super-$W_3^{cl}$:
\begin{eqnarray}
J^{cl}(Z_1)J^{cl}(Z_2) &=&\frac{\Tb\Db J^{cl}}{\Z}-
  \frac{\T\D J^{cl}}{\Z}
          +\frac{\T\Tb J^{cl}}{\Z^2}+\frac{\T\Tb\partial J^{cl}}{\Z}
     \nonumber \\
J^{cl}(Z_1)T^{cl}(Z_2) & = &
\frac{\Tb\Db T^{cl}}{\Z}-\frac{\T\D T^{cl}}{\Z}
+2\frac{\T\Tb T^{cl}}{\Z^2}
  +\frac{\T\Tb\partial T^{cl}}{\Z} \nonumber \\
 T^{cl}(Z_1) T^{cl}(Z_2) & = &
   \frac{\T\Tb}{\Z^2}U_3^{cl}+
  \frac{\T}{\Z} \Psi_{{7\over 2}}^{cl}
     -\frac{\Tb}{\Z} {\overline\Psi}_{{7\over 2}}^{cl}
 + \frac{\T\Tb}{2\Z}\left(
 \partial U_3^{cl} +\Db\Psi_{{7\over 2}}^{cl}
 +\D{\overline\Psi}_{{7\over 2}}^{cl} \right) , \label{9}
\end{eqnarray}
where the composite supercurrents are now given by
\begin{eqnarray}
\Psi_{{7\over 2}}^{cl} & = &   -72 T^{cl}\D J^{cl}
   + 8J^{cl}\D T^{cl} -128\left( J^{cl}\right)^2\D J^{cl} \nonumber \\
{\overline\Psi}_{{7\over 2}}^{cl} & = & -72 T^{cl}\Db J^{cl}
   + 8J^{cl}\Db T^{cl}   -128\left( J^{cl}\right)^2\Db J^{cl} \nonumber \\
 U_3^{cl} & = & 56J^{cl} T^{cl} +128\left( J^{cl}\right)^3 \; .\label{10}
\end{eqnarray}
These relations are guaranteed to define a closed nonlinear algebra (with all
the Jacobi identities satisfied) because they have been obtained from those of
$N=2$ super-$W_3$ algebra via a contraction procedure.

\setcounter{equation}{0}
\section{Free superfield realizations of super-$W_3$}

In this Section we construct two $N=2$ superfield realizations of
$N=2$ super-$W_3$ algebra. They prove to be closely related to each other.

The first realization is via two $N=2$ chiral spin $\frac{1}{2}$ fermionic
superfields $\chi(Z),\;\xi(Z)$,
\begin{equation}
\D\bar\chi = \Db\chi = 0 \quad , \quad \D\bar\xi = \Db\xi=0 \quad , \label{a2}
\end{equation}
with the two-point functions given by
\begin{equation}
\left\langle \chi (Z_1)\bar\chi (Z_2) \right\rangle = \frac{1}{\Z}
 + \frac{\T\Tb}{2\Z^2}
 \quad , \quad \left\langle \xi (Z_1)\bar\xi (Z_2) \right\rangle = \frac{1}{\Z}
 + \frac{\T\Tb}{2\Z^2}\;. \label{a6}
\end{equation}
We have explicitly checked (this is a rather tedious labor despite the fact
that we are using the condensed $N=2$ superfield formalism) that
the supercurrents
\begin{equation}\label{a5}
J=-\chi\bar\chi-\xi\bar\xi + \sqrt{\frac{c}{8}} \left( \Db\bar\chi
  - \D\chi \right)
\end{equation}
and
\begin{eqnarray}
 T & = &
\sqrt{\frac{c}{8}}\left( \partial\D\xi - \partial\Db{\bar\xi}\right)
 +2\partial\xi{\bar\chi}
+\partial\xi{\bar\xi}+\partial{\bar\xi}\xi-2\partial{\bar\xi}\chi+
\partial\chi{\bar\chi}-\partial{\bar\chi}\xi
-\frac{40}{c}\xi{\bar\xi}\chi{\bar\chi} \nonumber \\
 & + & \D\xi\left( \D\chi+\Db{\bar\chi}+\D\xi-3\Db{\bar\xi} \right) +
    \Db{\bar\xi}\left(- \D\chi-\Db{\bar\chi}+\Db{\bar\xi} \right)
	  \nonumber \\
 & + & {\sqrt{\frac{8}{c}}}\left[ \D\xi\left( 2\xi{\bar\chi}-
4{\bar\xi}\chi+\chi{\bar\chi}-\xi{\bar\xi}\right) +
\D\chi\left( \xi{\bar\chi}-2{\bar\xi}\chi-\xi{\bar\xi}\right) \right]
\nonumber \\
 & + & {\sqrt{\frac{8}{c}}}\left[ \Db{\bar\xi}\left(
-4\xi{\bar\chi}+2{\bar\xi}\chi+\chi{\bar\chi}-\xi{\bar\xi}\right)+
    \Db{\bar\chi}\left( 2\xi{\bar\chi}-{\bar\xi}\chi-\xi{\bar\xi}\right)
 \right] \label{a8}
\end{eqnarray}
obey the defining SOPE's (\ref{1}), (\ref{4}), (\ref{5}).

This realization naturally generalizes the one proposed in \cite{a4} for $N=2$
superconformal algebra (the corresponding supercurrent $J(Z)$ is the $\xi=0$
reduction of (\ref{a5})). The chiral superfields $\chi(Z)$ and $\xi(Z)$ are
recognized as supercurrents generating $N=2$ superextensions of two independent
complex $U(1)$
Kac-Moody algebras. Note the presence of the Feigin-Fuchs linear terms in
(\ref{a5}), (\ref{a8}). Just these terms ensure an
arbitrary central charge in the present case (at the classical level)
\footnote{In the quantum case $c$ is expected to be restricted to
discrete series by the unitarity reasonings \cite{aa}.}.

In order to obtain one more field realization of $N=2$ super-$W_3$, one
notices that
the SOPEs (\ref{a6}) can be reproduced starting from the following particular
representation of the supercurrents $\chi(Z)$, $\xi(Z)$
\begin{equation}
\chi = -\Db \Ub \quad , \quad \bar\chi = \D \U \quad , \quad
\xi = -\Db \Pb \quad , \quad \bar\xi = \D \Phi \; ,\label{a7}
\end{equation}
where ${\cal U}(Z)$, $\Phi(Z)$ are the spin 0 free chiral $N=2$ superfields
\begin{equation}
\D\Pb = \Db\Ph = 0 \quad ,\quad \D\Ub = \Db\U=0 \label{a1}
\end{equation}
\begin{equation}
\left\langle {\cal U} (Z_1)\Ub(Z_2) \right\rangle = \ln (\Z)
- \frac{\T\Tb}{2\Z}
\quad , \quad \left\langle \Phi (Z_1)\Pb (Z_2) \right\rangle =
 \ln (\Z) - \frac{\T\Tb}{2\Z} \;. \label{a4}
\end{equation}
To get the free-superfield expressions for $J(Z)$ and $T(Z)$ one should
replace the $U(1)$ supercurrents $\chi(Z)$, $\xi(Z)$ in eqs. (\ref{a5}),
(\ref{a8}) by their particular representation (\ref{a7}).
For brevity, we quote the free-superfield form only for the $N=2$
conformal supercurrent
\begin{equation}\label{a3}
J=\Db \Ub \D \U+ \Db \Pb \D \Phi -\sqrt{\frac{c}{8}}\partial (\U + \Ub )\quad .
\end{equation}

This realization generalizes the free chiral superfield realization
of $N=2$ superconformal algebra given in \cite{a3}. We conjecture that it
is closely related to the $N=2$ supersymmetric Toda system associated with the
superalgebra $sl(3|2)$\footnote{In a recent paper \cite{a14}
it has been observed that the extended classical symmetry of this system is
generated just by the spin 1 and spin 2 $N=2$ supercurrents.}.

Before ending this Section, it is worth saying a few words as to how we
arrived at the above particular realizations of $N=2$ super-$W_3$. These
were prompted to us while we treated this superalgebra in the framework of
the covariant reduction approach worked out earlier on
the simpler examples of $W_2$ (Virasoro) and $W_3$ algebras
\cite{{a15},{a1}}. Without
entering into details we only note that this approach allows to regard
nonlinear (super)algebras of the $W_N$ type as special realizations of some
associate linear higher-spin (super)algebras $W_{N}^{\infty}$. The
appropriate (super)currents and scalar (super)fields (e.g., of the type
considered above) naturally come out in the covariant reduction
approach as the
parameters of some coset (super)spaces connected with the
infinite-dimensional (super)algebras just mentioned. Then the covariant
relations between them of
the type (\ref{a5}), (\ref{a8}), (\ref{a3}) arise as a result of imposing
covariant constraints on the Cartan one-forms describing the
geometry of these cosets. In more detail the applications of this geometric
approach to $N=2$ super-$W_3$ will be reported elsewhere \cite{a16}. Here
we wish to stress that all the
formulas and statements of the present paper are self-contained in their own
right and do not require the reader to be familiar with the covariant
reduction techniques.

\setcounter{equation}0
\section{N=2 super Boussinesq equation}
In this Section we deduce $N=2$ super Boussinesq equation and give the
second hamiltonian structure for it.

It is well known that the bosonic Boussinesq equation has the second
hamiltonian structure which is equivalent to the classical form of the
$W_3$ algebra \cite{a17}, namely
\begin{equation}
{\dot T} =  \left[ T, H \right] \quad , \quad
{\dot W}  =  \left[ W, H \right] \quad , \label{b1}
\end{equation}
with
\begin{equation}
H=\int dz W \label{b2}
\end{equation}
and the currents $T(z)$ and $W(z)$ obeying the OPE's of the classical
$W_3$ algebra (with an arbitrary central charge)\footnote{The commutators
here and in the subsequent formulas are defined like in the quantum case
as the radially ordered OPE's (SOPE's).}.

It is natural to define $N=2$ super Boussinesq equation as the $N=2$
superfield equation the second hamiltonian structure for which is induced
by the $N=2$ super-$W_3$ algebra (\ref{1}), (\ref{4}), (\ref{5}). In other
words, we consider the set of the evolution equations
\begin{equation}\label{b3}
{\dot T}  =  \left[ T, H \right] \quad , \quad
{\dot J} =  \left[ J, H \right]
\end{equation}
where now hamiltonian $H$ is given by
\begin{equation}
H=\int dZ \left( T +\alpha J^2 \right) \; .
 \label{b4}
\end{equation}
We emphasize that the hamiltonian (\ref{b4}) is the most general one
which can be constructed out of $J$ and $T$ under
the natural assumptions that it respects $N=2$ supersymmetry and has the
same dimension 2 as the bosonic hamiltonian (\ref{b2}).
Note the appearance of the free parameter $\alpha$ in (\ref{b4}).

Now, using the previously established SOPE's (\ref{1}), (\ref{4}) and
(\ref{5}),
we immediately find the explicit form of the sought $N=2$ super Boussinesq
equation:
\begin{eqnarray}
{\dot T} & = & -2 J'''+\left[\Db ,\D\right]  T'+
 \frac{88}{c}\partial\left( \Db J \D J\right)
 -\frac{28}{c}J'\left[\Db , \D\right]J -\frac{12}{c}J\left[\Db , \D\right]J'
+\frac{256}{c^2}J^2J' \nonumber \\
 &+ & \left( \frac{40}{c}-2\alpha \right)\Db J\D  T
     +\left( \frac{40}{c}-2\alpha \right)\D J\Db  T
 +\left(\frac{64}{c}+4\alpha\right)J' T
  +\left(\frac{24}{c}+2\alpha\right)J T' \nonumber \\
{\dot J} & = & 2 T'+\alpha\left( \frac{c}{4}\left[\Db ,\D\right]J'
 +4JJ'\right) \quad .\label{b5}
\end{eqnarray}
Note that, in contrast to the set (\ref{b1}) which can be
equivalently rewritten as a single equation for the conformal stress tensor
$T$ (it is just what is usually called the Boussinesq equation), the set
(\ref{b5}) cannot be reduced to one equation for the conformal
supercurrent $J$. Thus the $N=2$ superextension of
Boussinesq equation in general amounts to the system of coupled equations
for 4 bosonic and 4 fermionic fields \footnote{It would be interesting to
compare eqs. (\ref{b5}) with another $N=2$ extension of Boussinesq equation
deduced in \cite{a7} within a generalized Lax
representation (using $N=1$ superfield formalism).}.

It is instructive to examine the bosonic subsector of (\ref{b5}), with all
fermions omitted:
\begin{eqnarray}
{\dot\phi} & = & 2v'+\frac{\alpha c}{4}u'+4\alpha\phi\phi' \nonumber \\
{\dot v}& = & -2\phi'''+\omega'-\frac{28}{c}\phi' u -\frac{12}{c}\phi u'
+\left(\frac{64}{c}+4\alpha\right)\phi'v+\left(\frac{24}{c}+2\alpha\right)
\phi v'+\frac{256}{c^2}\phi^2\phi' \nonumber \\
{\dot u} & = & 2\omega'+\frac{\alpha c}{4}\phi'''+4\alpha u \phi'+
4\alpha\phi u' \nonumber \\
{\dot\omega}& = & -2u'''+v'''-\frac{128}{c}uu'+\frac{60}{c}\phi'\phi''-
\frac{12}{c}\phi\phi'''+\frac{64}{c}uv'+\frac{512}{c^2}u\phi\phi'+
\frac{256}{c^2}\phi^2u' \nonumber \\
 &+& \left( 6\alpha+\frac{24}{c}\right)\omega\phi'+
\left(4\alpha+\frac{64}{c}\right)u'v+\left(2\alpha+\frac{24}{c}\right)
\phi\omega' \quad . \label{b6}
\end{eqnarray}
Here
$$ J|=\phi \quad ,\quad \left[ \Db ,\D\right] J| = u \quad , \quad
 T|=v \quad ,\quad \left[\Db ,\D\right] T| =\omega \quad .
$$

The set (\ref{b6}) contains the $N=0$ Boussinesq equation in a very
special manner. Namely, if we choose
\begin{equation}\label{b7}
\alpha =-\frac{4}{c}\;,
\end{equation}
then the equations (\ref{b6}) admit the following self-consistent truncation
\begin{equation}\label{b8}
 \phi = 0 \quad ,\quad u = 2v \;.
\end{equation}
In this case first of eqs. (\ref{b6}) is satisfied identically, while the
second and third ones turn out to coincide and, together with the fourth
equation, just give the Boussinesq equation
\begin{equation}\label{b9}
{\dot\omega}  =  -3v''' -\frac{288}{c}vv' \quad , \quad
{\dot v}  =  \omega '\quad .
\end{equation}

Finally, we mention that the spinor and scalar $N=2$ superfield realizations
found for $J(Z)$ and $T(Z)$
in the previous Section give generalized
super-Miura maps for the $N=2$ Boussinesq equation (\ref{b5}). The explicit
form of the evolution equations for the Kac-Moody supercurrents
$\xi$, $\chi$
\begin{equation}
\dot{\xi} = \left[ H, \xi \right], \;\; \dot{\chi} = \left[ H, \chi \right]
\label{b10}
\end{equation}
as well as for the scalar superfields ${\cal U}$, $\Phi$ can be
easily established expressing the hamiltonian in terms of these superfields and
further employing the relations (\ref{a6}), (\ref{a4}). Eqs. (\ref{b10}) are
related to eqs. (\ref{b5}) like the mKdV equation to the KdV one.

We postpone the analysis of the integrability properties of the $N=2$ super
Boussinesq equation (the existence of the Lax pair and infinite series of
the conserved quantities, etc) to future publications.

\setcounter{equation}0
\section{Conclusion}
To summarize, we have concisely rewritten classical $N=2$ super-$W_3$
algebra of \cite{a6} in terms of two $N=2$ supercurrents, found its
Feigin-Fuchs type representations, in terms of two chiral $N=2$ $U^c(1)$
Kac-Moody supercurrents and two free scalar chiral $N=2$ superfields, and
constructed a one-parameter family of $N=2$ super Boussinesq equations
the second hamiltonian structure for which is related to this superalgebra.
We have also deduced a new classical nonlinear superalgebra
$N=2$ super-$W_3^{cl}$ by taking the contraction limit $c=0$ in the defining
relations of $N=2$ super-$W_3$. In a forthcoming publication we will
extend our consideration to the case of full quantum $N=2$ super-$W_3$
algebra of ref. \cite{aa}.

It is interesting to apply our manifestly $N=2$ supersymmetric
superfield formalism  for constructing higher-$N$
superextensions of $W_3$. For instance, $N=2$ superconformal algebra can be
extended to the $N=4$ $SU(2)$ one by adding a spin 1 chiral $N=2$
supercurrent to $J(Z)$ (this additional supercurrent should possess the
$U(1)$ charge $+2$, if one ascribes the charge $+1$ to the $N=2$ superspace
spinor coordinate $\theta$). To preserve the algebraic structure, we also
have to add some extra $N=2$ supercurrents to $T(Z)$ in order
to complete the latter to an irreducible $N=4$ supermultiplet (it can be
primary or quasi-primary with respect to the $N=4$ superconformal algebra)
and then try to write a closed set of SOPE's between all these supercurrents.
The minimal way to enlarge $T(Z)$ so as to have still
only one spin 3 current is as follows: we should add one extra real spin 1
$N=2$ supercurrent and one complex spin $3\ /2$ $N=2$ supercurrent (besides
the chiral supercurrent extending $N=2$ superconformal algebra to the $N=4$
one). It is an open question whether these supercurrents can be forced to
generate an $N=4$ super-$W_3$ algebra. \vspace{0.5cm}

\noindent ACKNOWLEDGMENTS

We would like to thank Professor Abdus Salam, the International Atomic
Energy Agency and UNESCO for hospitality at the International Centre for
Theoretical Physics, Trieste, where this work has been completed.
We are also grateful to Dr. R. Malik for many useful and clarifying
discussions.

\end{document}